\documentclass[twocolumn,secnumarabic,amssymb, nobibnotes, aps, prd, article]{revtex4-2}
\usepackage[squaren]{SIunits}
\usepackage{graphicx}
\usepackage{amsmath}

\usepackage{graphicx}
\usepackage{dcolumn}
\usepackage{bm}
\usepackage{physics}
\usepackage{amsmath}
\usepackage{natbib}
\usepackage{booktabs}
\usepackage{array}
\usepackage{float}
\usepackage{multirow}
\usepackage[colorlinks,linkcolor=blue,urlcolor=blue,anchorcolor=blue,citecolor=blue]{hyperref}
\usepackage{placeins}

\begin{document}

\title{Spin texture and tunneling magnetoresistance in atomically thin CrSBr}%

    \author{Ziqi Liu$^{1}$}
    \author{Yichang Sun$^{1}$}
    \author{Chengfeng Zhu$^{1}$}
    \author{Canyu Hong$^{2}$}
    \author{Yuchen Gao$^{1}$}
    \author{Zeyuan Sun$^{2}$}
    \author{Kenji Watanabe$^{3}$}
    \author{Takashi Taniguchi$^{4}$}
    \author{Shiwei Wu$^{2}$}
    \author{Zuxin Chen$^{5}$}
    \email{chenzuxin@m.scnu.edu.cn}
    \author{Pingfan Gu$^{6}$}
    \email{gupingfan@njust.edu.cn}
    \author{Yu Ye$^{1,7,8}$}
    \email{ye\_yu@pku.edu.cn}
    
\affiliation{
$^{1}$State Key Laboratory for Artificial Microstructure $\rm{\&}$ Mesoscopic Physics and Frontiers Science Center for Nano-optoelectronics, School of Physics, Peking University, Beijing, 100871, China\\
$^{2}$State Key Laboratory of Surface Physics, Key Laboratory of Micro and Nano Photonic Structures (MOE), and Departement of Physics, Fudan University, Shanghai 200433, China\\
$^{3}$Research Center for Electronic and Optical Materials National Institute for Materials Science 1-1 Namiki, Tsukuba 305-0044, Japan\\
$^{4}$Research Center for Materials Nanoarchitectonics National Institute for Materials Science 1-1 Namiki, Tsukuba 304-0044, Japan\\
$^{5}$School of Semiconductor Science and Technology, South China Normal University, Foshan 528225, China\\
$^{6}$MIIT Key Laboratory of Semiconductor Microstructure and Quantum Sensing Department of Applied Physics, Nanjing University of Science and
Technology, Nanjing 210094, China\\
$^{7}$Collaborative Innovation Center of Quantum Matter, Beijing 100871, China\\
$^{8}$Liaoning Academy of Materials, Shenyang, 110167, China\\
}
\date{\today}

\begin{abstract}
The exploration of spin configurations and magnetoresistance in van der Waals magnetic semiconductors, particularly in the realm of thin-layer structures, is of paramount significance for the development of two-dimensional spintronic nanodevices. In this letter, we present detailed magneto-transport and photoluminescence studies on few-layer CrSBr flakes utilizing a vertical tunneling device configuration. Our investigations revealed complex magnetic states along the evolutionary path of few-layer CrSBr. We observed intermediate states exhibiting identical net magnetization demonstrate rectification properties, reminiscent of a diode-like behavior at positive and negative bias voltages. Notably, in devices with 5-layer CrSBr, we detected an intriguing positive magnetoresistive state when subjected to an in-plane magnetic field along the $b$-axis. The implementation of the Mott two-current model successfully calculated the tunneling resistance of different magnetic states, thereby elucidating the spin configurations responsible for the observed transport phenomena. These insights not only provide new perspectives on the intricate spin textures of two-dimensional CrSBr but also highlight the efficacy of tunneling measurements as a sensitive method for probing magnetic order in van der Waals materials.
\end{abstract}
\maketitle


\section{Introduction}

Magnetic semiconductors play a crucial role in the field of condensed matter physics and materials science owing to their unique combination of semiconducting and magnetic properties \cite{dietl_ten-year_2010, wang_very_2018, huang_layer_2017, gong_discovery_2017}. This amalgamation presents promising opportunities for advances in spintronics, magnetic storage, and quantum computing \cite{huang_electrical_2018, jiang_controlling_2018, verzhbitskiy_controlling_2020, wang_magnetic_2022}. The ability to manipulate both electrical current flow and magnetic moment orientation within a single material has great potential to improve the efficiency and versatility of electronic devices, with potential applications in energy-efficient electronics \cite{krempasky_efficient_2023, zhang_highly_2021, gu_magnetic_2022, gu_multi-state_2023}. However, the controlled and stable realization of the desired combination of semiconducting and magnetic properties in materials remains a formidable challenge. 

The recently discovered van der Waals (vdW) material CrSBr has garnered significant attention as a highly promising candidate due to its distinctive direct bandgap semiconductor characteristics and layered A-type antiferromagnetic (AFM) order \cite{goser_magnetic_1990}. Notably, CrSBr exhibits a pronounced spin-charge interaction, where its band structure exhibits a strong dependence on its magnetic order through interlayer interactions \cite{wilson_interlayer_2021}. Consequently, obtaining a comprehensive understanding of the evolution of spin textures in atomically thin CrSBr under external magnetic fields is of utmost importance for understanding the spin-charge interaction in a $d$-band hosting magnetic semiconductor, as well as for designing vdW AFM-based spintronic devices. Experimental investigations have demonstrated that below the Néel temperature of 132 K, CrSBr adopts an A-type AFM state \cite{goser_magnetic_1990}. Further studies have unveiled an intermediate ferromagnetic (iFM) state above the Néel temperature, characterized by intralayer ferromagnetic (FM) order and random interlayer magnetic order along the $c$-axis, with the critical temperature for intralayer magnetization ($T_{\rm{C}}^{\rm{intra}}$) decreasing from $\sim$160 K in bulk to $\sim$146 K in monolayers \cite{lee_magnetic_2021, telford_layered_2020}. Moreover, zero-field muon spin rotation (ZF-$\mu$SR) studies have indicated a deceleration of magnetic fluctuations below $\sim$100 K, with a spin freezing process occurring around 40 K \cite{lopez_dynamic_2022}. Electrical transport and specific heat measurements have identified two additional magnetic phase transitions at higher temperatures (156 K and 185 K) \cite{liu_three_2022}, underscoring the intricate magnetic nature of CrSBr. Previous studies have reported layer-by-layer spin flips and overall spin-flop processes in few-layer CrSBr under an external magnetic field along the $b$-axis \cite{ye_layer-dependent_2022, wilson_interlayer_2021, sun_resolving_2025}. However, the detailed mechanisms of each magnetic phase transition still remain unclear based on the present experimental data. 

Within the domain of vdW magnetic materials, especially AFM semiconductors or insulators, the absence of a discernible magnetic moment renders tunneling magnetoresistance an exceptionally effective and sensitive method for characterizing their magnetic properties. \cite{dong_tunneling_2022, klein_probing_2018}. Tunneling probability is found to be exponentially related to both the tunneling height and width. It has been demonstrated that even slight changes in these factors can lead to significant modifications in tunneling probability \cite{yang_macroscopic_2024, boix-constant_multistep_2023}. Consequently, variations in the band structure and changes in material thickness, which occur during magnetic phase transitions, can significantly influence the tunneling probability, thereby affecting the tunneling resistance \cite{wang_very_2018, gu_multi-state_2023, ye_layer-dependent_2022}. Furthermore, the tunneling process is capable of distinguishing a broader range of magnetic states and providing more detailed insights into their characteristics, such as their spatial symmetry, since the tunneling barrier is influenced by both the bias voltage and the magnetic structure in real space\cite{fu_tunneling_2024}. Given that the tunneling current is obtained by integrating the tunneling probability within the energy range that spans the Fermi levels of two metal contacts, thermal excitation has minimal impact on it \cite{heath_role_2020}.

This study explores the spin configurations and evolution of few-layer CrSBr under external magnetic fields using vertical tunneling devices. When subjected to an in-plane magnetic field along the $b$-axis, CrSBr predominantly adopts collinear magnetic states, with a distinct spin-flip evolution observed during the field sweep. The combination of photoluminescence (PL) spectroscopy and tunneling magnetoresistance measurements unveils a layer-resolved magnetic evolution path and magnetic states with identical net magnetization but different tunneling resistance. Remarkably, in 5-layer CrSBr devices, a positive magnetoresistive state emerges when an in-plane field is applied along the $b$-axis. A two-channel spin current model is developed to calculate the tunneling current and interpret the experimental results. In summary, the present findings offer original insights into the intricate spin structure of CrSBr and underscore the sensitivity of tunneling measurements as a potent technique for probing magnetic order in vdW materials.

\begin{figure*}[ht!]
\centering 
\includegraphics[width=1\textwidth]{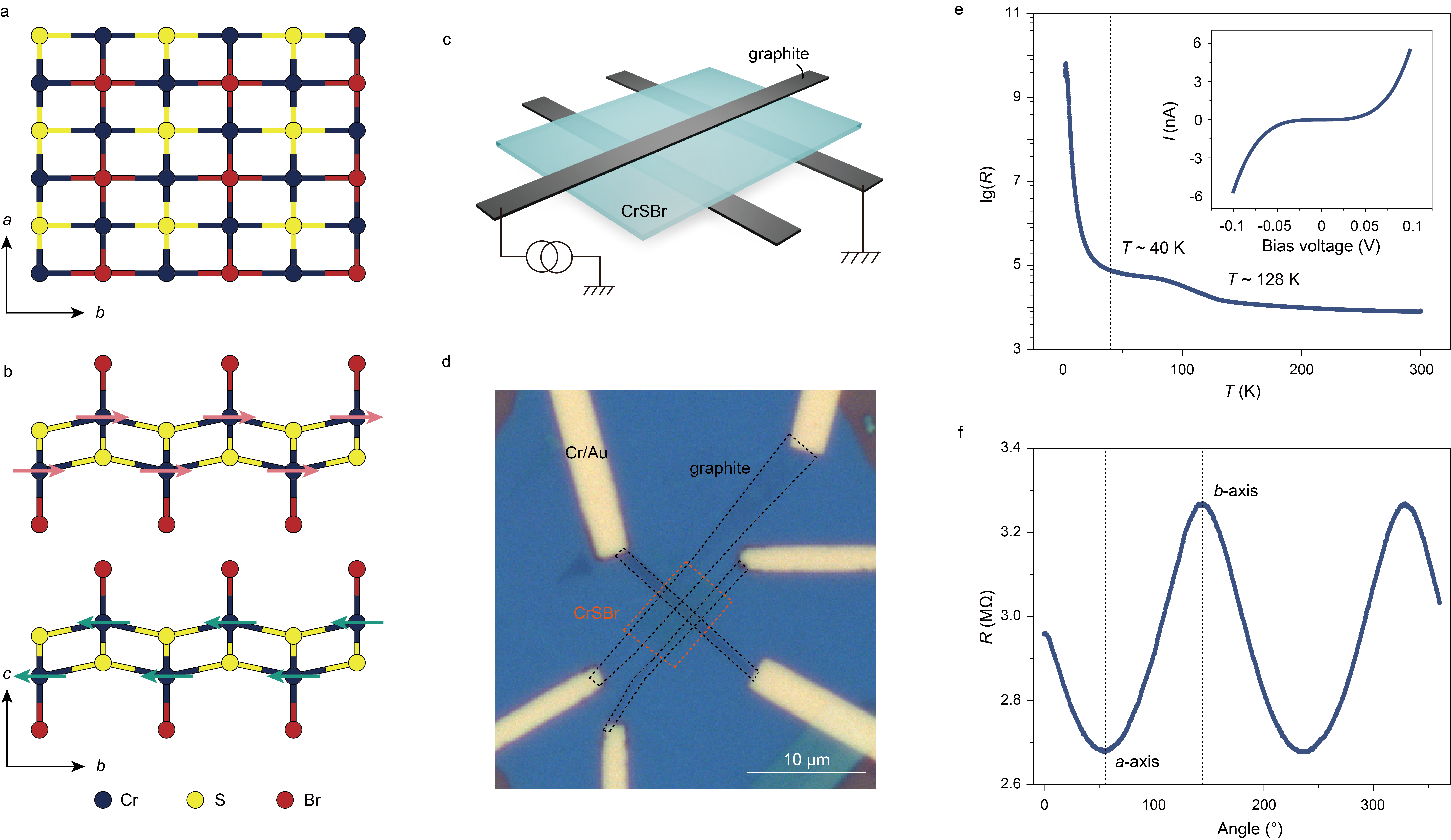}
\caption{\textbf{Device structure and basic characterizations.} 
\textbf{(a),} Top view of the atomic structure of CrSBr. \textbf{(b),} Side view of the CrSBr atomic structure with arrows indicating the spin orientation of each Cr atom, revealing an A-type AFM ground state. \textbf{(c, d),} Schematic and optical image of the device. \textbf{(e),} Tunneling resistance of a 4-layer CrSBr device from 300 K to 2 K at a bias voltage of 10 mV. The dashed lines in the figure mark the two inflection points at 40 K and 128 K. The inset shows the $I$-$V$ curve of the device at 2 K. \textbf{(f),} Tunneling resistance of the device with an in-plane magnetic field. The angles of the magnetic field corresponding to the $a$- and $b$-axes are labeled in the figure. The bias voltage is 80 mV, and the magnitude of the magnetic field is 1 kG.
}
\label{F1}
\end{figure*}

\section{Results and Discussion}
Single crystals of CrSBr are synthesized using the chemical vapor transport (CVT) technique and subsequently exfoliated into few-layer flakes using the conventional scotch tape method (see Methods). CrSBr possesses an orthorhombic lattice with cell dimensions of $a$=3.50 ${\angstrom}$, $b$=4.76 ${\angstrom}$, and $c$=7.96 ${\angstrom}$ \cite{telford_layered_2020}, where each CrSBr layer is stacked along the $c$-axis (Fig. \ref{F1}a, b). Notably, CrSBr exhibits intralayer ferromagnetism along the $b$-axis, while interlayer A-type AFM coupling occurs along the $c$-axis (Fig. \ref{F1}b). To investigate the magnetic transitions of CrSBr, we fabricated vertical tunneling devices based on few-layer CrSBr flakes (Fig. \ref{F1}c). Cross-structured few-layer graphite stripes are employed to establish contact with the CrSBr tunneling layer, and the entire device is encapsulated by hexagonal boron nitride (\textit{h}-BN, see Methods). The intentionally narrow width of the graphite layer aims to minimize the area of the tunneling junction, thereby reducing the impact of multi-domain effects in the junction area (\cite{boix-constant_multistep_2023, ye_layer-dependent_2022}; for detailed confirmation of domain size, see Supplementary Fig. S1\cite{supplemental_material})\nocite{xie_engineering_2023, yu_direct_2024,tschudin_imaging_2024}. The number of CrSBr layers is determined through atomic force microscope measurements (Supplementary Fig. S2)\cite{supplemental_material}.

Initially, the resistance of the devices was measured over a temperature range of 300 K to 2 K. Due to the semiconducting nature of CrSBr, the devices exhibited a resistance in the range of a few thousand ohms at room temperature (Fig. \ref{F1}e). However, contrary to expectations based on typical semiconductor behavior, the device resistance did not show an exponential increase as the temperature decreased. Instead, a minor kink was observed at the Néel temperature ($\sim$128 K), indicating the emergence of AFM magnetic order \cite{lee_magnetic_2021, telford_coupling_2022, telford_layered_2020}. Notably, the device resistance displayed a significant increase as the temperature decreased below 40 K, suggesting the dominance of the tunneling process (inset of Fig. \ref{F1}e). This behavior may be associated with the spin-freezing process below 40 K \cite{lopez_dynamic_2022}. Due to the strong in-plane anisotropy of CrSBr, the tunneling resistance exhibited precise 180$^{\degree}$ periodic symmetry when a small external magnetic field was applied and rotated in the $ab$-plane (Fig. \ref{F1}f), providing a means to determine the orientation of the $a$- and $b$-axes (as confirmed by polarized Raman spectra in Supplementary Fig. S2d)\cite{supplemental_material}. 

Following the determination of the $a$- and $b$-axes, the tunneling resistance was measured by sweeping a magnetic field along the $a$- or $b$-axis at a fixed bias voltage to investigate the magnetic behavior of the few-layer CrSBr. The magnetoresistance (MR) was calculated using the formula:

\vspace{-2mm}
\begin{equation}
{\rm{MR}} = \frac{R(B)-R(B=0)}{R(B=0)}\times100\%
\end{equation}

\begin{figure}[tb]
	\includegraphics[width=1\columnwidth]{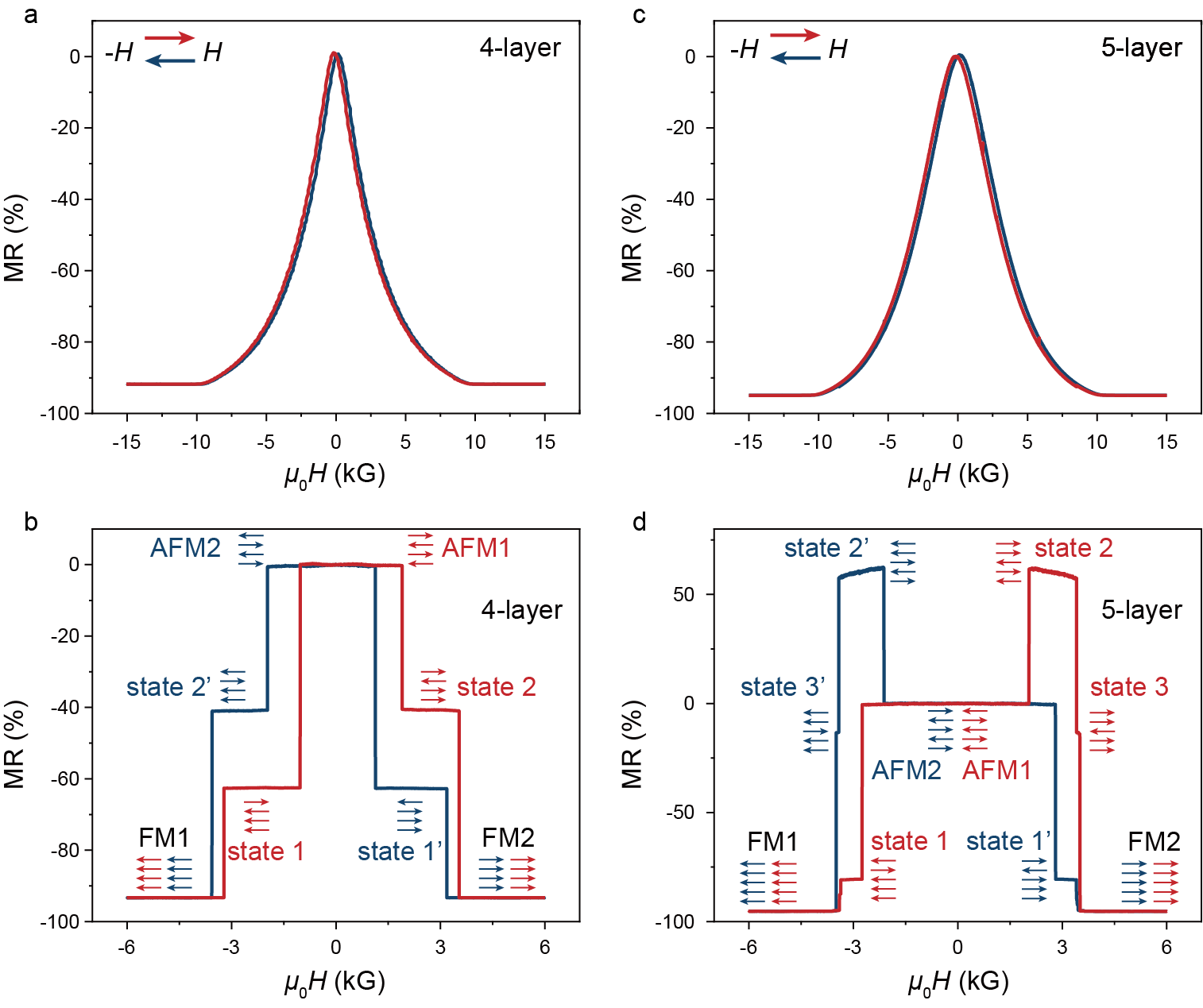}
	\caption{\label{F2} \textbf{Layer-dependent tunneling MR with magnetic field sweeping along the $a$- and $b$-axes.} 
\textbf{(a),} Tunneling resistance of a 4-layer CrSBr device with external magnetic field sweeping back and forth along the $a$-axis at 2 K and 70 mV bias voltage. \textbf{(b),} Tunneling resistance of the 4-layer CrSBr with the external magnetic field sweeping along the $b$-axis at 60 mV bias voltage. Two intermediate states with different resistance values are observed. \textbf{(c),} Tunneling resistance of a 5-layer CrSBr with the external magnetic field sweeping along the $a$-axis at 100 mV basis voltage.  \textbf{(d),} Tunneling resistance of the 5-layer CrSBr with the external magnetic field sweeping along the $b$-axis at 100 mV bias voltage. A positive magnetoresistive state with an MR of 50\% is observed. The spin configuration of each state is marked in the graphs.}
\end{figure}

When a magnetic field was applied along the $a$-axis, the 4- and 5-layer CrSBr devices exhibited consistent behavior. Specifically, the tunneling resistance decreases with increasing magnetic field and eventually saturates, indicating a negative MR (Fig. \ref{F2}a, c). This observation aligns with the magnetic ground state of CrSBr, which is magnetized along the $b$-axis and features interlayer A-type AFM coupling. In the presence of an applied magnetic field, the magnetization of the layers gradually tilts towards the $a$-axis and eventually becomes fully magnetized along the $a$-axis \cite{wilson_interlayer_2021}. In the FM state, the spin-up or spin-down electrons encounter lower tunneling barriers due to the uniform direction of magnetization in all CrSBr layers, resulting in significantly lower tunneling resistance than in the AFM state.

Nevertheless, when a magnetic field is applied along the $b$-axis, the 4- and 5-layer devices display distinct transport properties, with each device undergoing a multistep spin-flip process. These transitions are accompanied by sharp changes in tunneling resistance (Figs. \ref{F2}b and d). In the case of the 4-layer CrSBr device, the material undergoes a transition from an A-type AFM state (high-resistance state) to an FM state (low-resistance state), resulting in a negative MR. Between the AFM and FM states, the tunneling resistance assumes an intermediate value, manifesting substantial hysteresis. Namely, under identical bias voltage and magnetic field conditions, the tunneling resistance of the device manifests two discrete resistance values during the ascending-field and descending-field sweeps. Similarly, the tunneling resistance displays two different values, low (high) and high (low), under negative and positive magnetic fields in this intermediate state, respectively, during the ascending-field (descending-field) sweep (Fig. \ref{F2}b). This behavior suggests the presence of two different intermediate magnetic states.

\begin{figure*}[ht!]
\centering 
\includegraphics[width=1\textwidth]{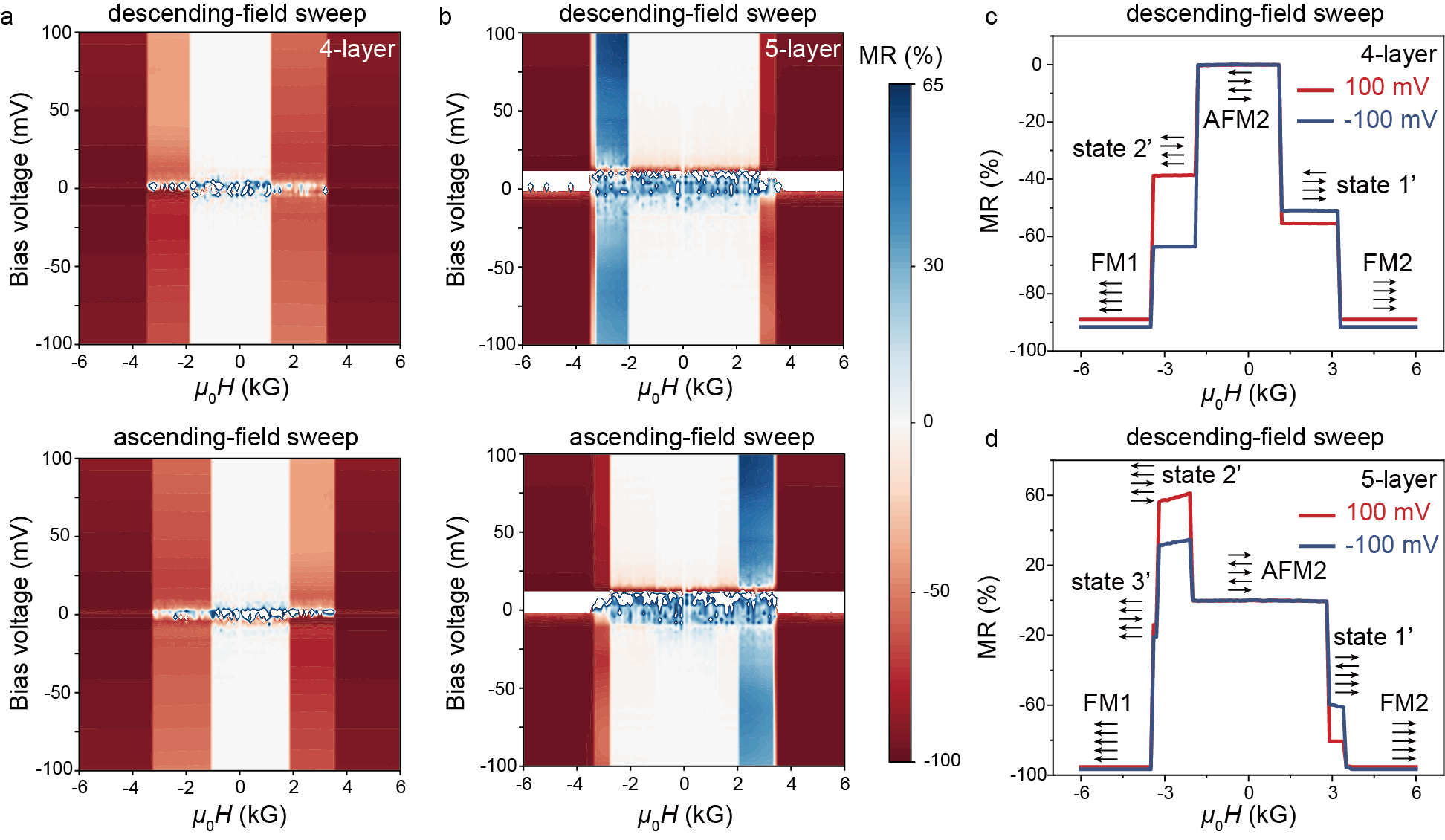}
\caption{\textbf{Rectification properties of asymmetric spin configurations.} 
\textbf{(a),} $I$-$V$ curve mapping of a 4-layer CrSBr device under the external magnetic field along the $b$-axis descending-field (upper panel) and ascending-field (lower panel) sweeps. \textbf{(b),} $I$-$V$ curve mapping of a 5-layer CrSBr device under the external magnetic field along the $b$-axis descending-field (upper panel) and ascending-field (lower panel) sweeps. MR values near zero bias voltage are noisy due to the measurement resolution limit. \textbf{(c,d),} MR line-cuts of 4-layer and 5-layer CrSBr devices in panels \textbf{(a)} and \textbf{(b)} in the descending-field sweeps at $\pm$ 100 mV bias voltages.
}
\label{F3}
\end{figure*}

In the case of the 5-layer device, a total of five distinct tunneling magnetoresistive states are observed during the evolution under a magnetic field along the $b$-axis. Taking the ascending-field sweep as an illustrative example, the device transitions from the low-resistance state of the FM phase (FM1) to a spin-flip intermediate resistive state (state 1) and then to the high-resistance state of the AFM phase (AFM1) during the field's ascent. As the positive magnetic field increases, the device enters another magnetic state (state 2) with even higher resistance than the AFM state, exhibiting a positive MR of up to 50\%. Subsequently, it undergoes another spin-flip intermediate resistance state (state 3) to reach the FM low-resistance state (FM2) polarized along the positive magnetic field. It should be noted that the states in descending field process are time-reversal states in the ascending field process (Fig. \ref{F2}d).

To further substantiate the evolution of the magnetic states, we conducted measurements of the $I$-$V$ curves while sweeping the magnetic field (Figs. \ref{F3}a and \ref{F3}b). Of note, certain specific magnetic configurations exhibit distinct rectification properties. For instance, in the descending magnetic field sweep of the 4-layer CrSBr device, two intermediate states display contrasting diode-like rectification characteristics, namely, state 1$^{\prime}$ (2$^{\prime}$) demonstrating smaller MR under positive (negative) bias voltages (Fig. \ref{F3}c). Similarly, in the 5-layer CrSBr device, states 1$^{\prime}$ and 2$^{\prime}$ exhibit different tunneling MR under positive and negative bias voltages, while state 3$^{\prime}$ demonstrates nearly identical MR (Fig. \ref{F3}d). These findings suggest that in the 5-layer CrSBr devices, the magnetic configurations of states 1$^{\prime}$ and 2$^{\prime}$ exhibit broken symmetry in real space, whereas the magnetic configuration of state 3$^{\prime}$ is spatially inversion symmetric. Importantly, this rectification behavior is determined by the spin configurations governed by the magnetization history, rather than being influenced by the voltage sweep history (Supplementary Fig. S3)\cite{supplemental_material}. Additionally, certain asymmetric resistances under positive and negative bias voltages observed in the 4- and 5-layer CrSBr devices remain constant regardless of the magnetization history, potentially due to device asymmetry or discrepancies between the two graphite electrodes.

Notably, in the 4-layer CrSBr device, four distinct intermediate state tunneling MR values can be observed under positive and negative voltages (Fig. \ref{F3}c), indicating that states 1$^{\prime}$ and 2$^{\prime}$ are not entirely equivalent under opposite bias polarities. These findings further suggest that these two intermediate states possess different spatial symmetries, implying a divergent magnetic evolution pathway in our sample compared to the 4-layer CrSBr depicted in previous PL results \cite{wilson_interlayer_2021}. Therefore, we conducted PL spectroscopy on the same 4-layer sample previously examined for tunneling resistance to elucidate the layer-specific magnetic reversal pathways. Apart from the FM and  AFM state spectra, we noted two distinct PL outcomes for state 1 (1$^{\prime}$) and state 2 (2$^{\prime}$) (Fig. \ref{F4}a), indicating that state 1 and state 2 are not mirror-symmetric to each other. That is to say, states 1 (1$^{\prime}$) and 2 (2$^{\prime}$) exhibit different numbers of interlayer AFM- and FM-coupled interfaces, resulting in distinct PL spectra. Similar phenomena were also observed in the PL spectra of the 3-layer CrSBr previously\cite{tabataba-vakili_doping-control_2024}. Furthermore, the PL spectrum of state 1 (1$^{\prime}$) bears a resemblance to those of the FM states, while the spectrum of state 2 (2$^{\prime}$) demonstrates a greater similarity to those of the AFM states, both in terms of peak shape and position (Fig. \ref{F4}a). This alignment suggests that state 1 (1$^{\prime}$) contains a greater number of FM interfaces, and state 2 (2$^{\prime}$) contains a greater number of AFM interfaces.

On the basis of tunneling magnetoresistance measurements and PL spectroscopy results, we have established the magnetic evolutionary routes for 4- and 5-layer CrSBr (Figs. \ref{F4}b and \ref{F4}c). Here, we take the ascending-field sweep of the 4-layer CrSBr as an example (Figs. \ref{F4}b), when subjected to a substantially large negative external magnetic field, the spin orientation of each layer aligns with the external field, exhibiting an FM (FM1, $\downarrow\downarrow\downarrow\downarrow$) state along the $b$-axis. To ensure more FM interfaces, one of the two surface layers undergoes a spin flip as an initial step as the external field increases. Due to the device's asymmetric character, one of these two layers is more likely to undergo the initial flip, reaching state 1 (e.g., $\uparrow\downarrow\downarrow\downarrow$). With a further increase of the external field, one middle layer also flips its spin orientation, causing CrSBr to reach the A-type AFM ground state (AFM1, $\uparrow\downarrow\uparrow\downarrow$). As the magnetic field sweeps towards positive values, the surface layer that is antiparallel to the magnetic field flips, transitioning CrSBr into state 2 ($\uparrow\downarrow\uparrow\uparrow$). Subsequently, the spin orientation of the surface layer undergoes a flip, resulting in CrSBr transforming into an FM state aligned with the external field (FM2, $\uparrow\uparrow\uparrow\uparrow$). The descending-field sweep represents a time-reversal operation of the ascending-field sweep, during which the 4-layer CrSBr goes through state 1$^{\prime}$ and state 2$^{\prime}$ (Fig. \ref{F4}b).

Similarly, in the case of a 5-layer CrSBr device during an ascending-field sweep, the initial FM state transitions to the AFM ground state through two sequential spin-flip processes. For the odd-layer samples, the uncompensated layer's net magnetization in the AFM state breaks the energy degeneracy under a non-zero external magnetic field. As the positive external magnetic field increases, the spin orientations of both surface layers align in the opposite direction to the external field. With further increase of the magnetic field, the spins of the two surface layers flip one after the other, followed by the spin of the intermediate layer, leading to the sample transitioning into the FM state, magnetized according to the direction of the magnetic field. The descending-field sweep represents the time-reversal process of the ascending-field sweep (Fig. \ref{F4}c).

Interestingly, in the 4-layer CrSBr device, although state 1 possesses the same net magnetization as state 2, it is not the energetically preferred state under straightforward analysis. This is attributed to state 1 having more FM-coupled interfaces. During the spin-flip process transitioning from the FM1 state to state 1, if we assume uniform AFM coupling energies between each adjacent layer, the energy determinant is the second or third layer rather than the top layer flipping (Fig. \ref{F4}b). However, CrSBr is susceptible to external strain \cite{cenker_reversible_2022}, which could be randomly introduced during exfoliation or transfer processes. The markedly different PL spectra observed in state 1, along with tunneling resistance, confirm its magnetic configuration to be $\uparrow\downarrow\downarrow\downarrow$ rather than $\downarrow\uparrow\downarrow\downarrow$. Furthermore, we conducted an analysis of device discrepancies across three 4-layer CrSBr devices, revealing consistent step-like transitions in all devices, as illustrated in Supplementary Fig. S4\cite{supplemental_material}. Minor deviations in transition fields and MR values may stem from slight variations in interlayer coupling strength and layer-specific spin filter efficacy introduced during the fabrication process. A more detailed discussion on the magnetic evolution of CrSBr can be found in Supplementary Note S2\cite{supplemental_material}.

\begin{figure}[tb]
\includegraphics[width=1\columnwidth]{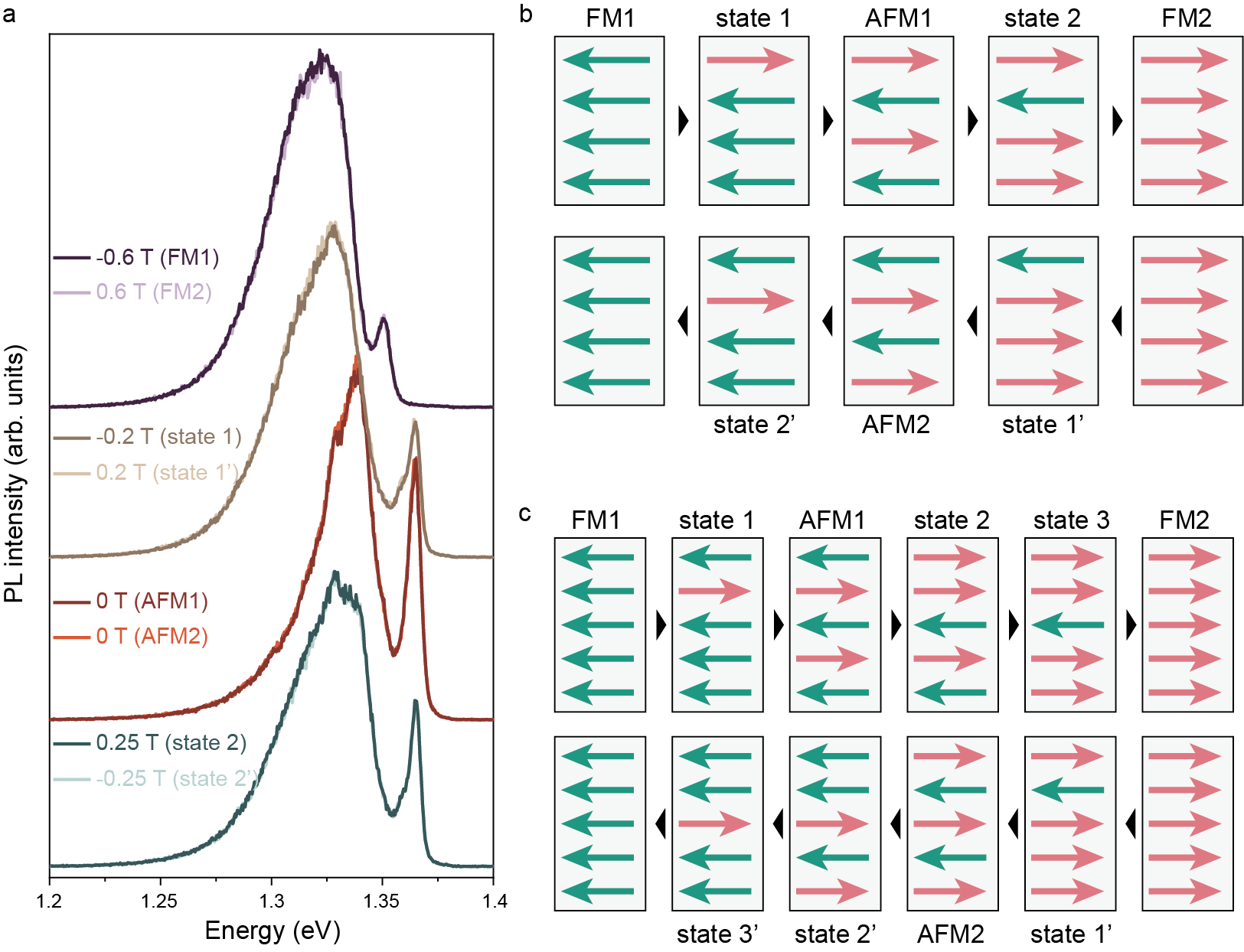}
\caption{\label{F4} \textbf{PL spectroscopy and the magnetic reversal route.} 
\textbf{(a),} PL spectra of different magnetic states in a 4-layer CrSBr device. Each line on the graph is annotated with the corresponding magnetic states and external field conditions. The time-reversal states exhibit identical PL peaks, while states 1 and 2 display distinct PL spectra. \textbf{(b),} Schematic diagram illustrating the evolution of the spin configuration of the 4-layer CrSBr with the magnetic field. The green and orange arrows indicate the spin direction of each layer of CrSBr. The black triangles indicate the ascending and descending field processes. \textbf{(c),} Schematic diagram illustrating the evolution of the spin configuration of the 5-layer CrSBr with the magnetic field. 
}
\label{F4}
\end{figure}

Subsequently, we employed the Mott two-current tunneling model to compute the MR of various magnetic states along the evolutionary path. In the tunneling process involving collinear magnetic states, spin-flip scattering is neglected since most scattering events conserve spin, enabling two distinct channels for electrons tunneling through CrSBr layers. The total tunneling current is then the aggregate of currents from spin-up and spin-down electrons. Under this framework, we posit that the tunneling current is directly proportional to the probability of an electron tunneling through the CrSBr layers. Noteworthy is the model's revelation that state 1 and state 2 exhibit markedly different tunneling resistances. Additionally, the tunneling barrier on each CrSBr layer is modulated by the presence of a bias voltage, affecting the spin-filter effect of different layers and resulting in divergent tunneling resistances between these states (Supplementary Note S1) and a simplified representation of the tunneling barriers of two-spin channel model is provided in Supplementary Figure S5\cite{supplemental_material}. State 1$^{\prime}$ (state 2$^{\prime}$) represents the time-reversed equivalent of state 1 (state 2), and as per the two-spin channel model, it is anticipated to possess the same tunneling resistance as state 1 (state 2), a finding corroborated by experimental data (Fig. \ref{F2}b). Crucially, despite state 1 and state 2 sharing identical net magnetization, they are distinguishable based on their tunneling resistance and PL spectra. In essence, reversing the current direction in these states would induce a change in their tunneling resistance, leading to rectification of the $I$-$V$ curve.

However, while the spin configurations during the field sweep have been determined, the underlying cause of positive MR in the $\downarrow\downarrow\uparrow\downarrow\uparrow$ ($\uparrow\uparrow\downarrow\uparrow\downarrow$) magnetic state remains unknown. The two-spin channel model provides unconventional insights, suggesting a more complex origin. First, there is a one-to-one correspondence between the spin configuration and the band structure in this material, with certain spin configurations potentially having larger energy band gaps, leading to increased tunneling barriers \cite{long2020persistence}. Secondly, the charge transfer resulting from the difference in work functions between CrSBr and graphite may influence the spin-filtering effect of CrSBr in close proximity to the graphite layer, resulting in positive MR. Additionally, the proximity effect \cite{ghiasi_electrical_2021} caused by the strong spin-orbit coupling of CrSBr in graphite may also impact the MR of the tunneling junction. While the precise underlying mechanisms require further elucidation, the distinct spin-filter effects induced by these phenomena in disparate layers have the potential to yield positive MR (Supplementary Note. S2)\cite{supplemental_material}. Due to the asymmetry of the spin configurations, reversing the current direction in the $\downarrow\uparrow\downarrow\downarrow\downarrow$ and $\uparrow\uparrow\downarrow\uparrow\downarrow$ states leads to two additional tunneling resistances (Fig. \ref{F3}d).

In conclusion, our study investigated the magnetic properties and spin textures of few-layer CrSBr. We identified distinct spin configurations that evolve with an in-plane magnetic field applied along different crystal axes. Notably, we observed some intermediate states that exhibited rectification behaviors under positive and negative biases. Intriguingly, a positive magnetoresistive state emerged in 5-layer CrSBr devices with a magnetic field along the $b$-axis. To gain insights into these phenomena, we developed a two-current model to compute the MR values, which helped to elucidate the underlying spin textures responsible for the observed transport characteristics and PL spectra. Our findings provide novel insights into the complex spin structure of CrSBr, revealing subtle spin textures on the evolution route. Furthermore, our work underscores the potential of tunneling magnetoresistance as a sensitive probe for investigating magnetic order in 2D materials. Further investigations focusing on spin dynamics and proximity effects in CrSBr heterostructures may unveil additional opportunities for spintronic applications.

\section{Methods}
\vspace{3mm}

\noindent \small{\textbf{Crystal synthesis.}
CrSBr single crystals were synthesized using the CVT method. Disulfur dibromide and chromium metal were mixed in a molar ratio of 7:13, and the mixture was sealed in a silica tube under vacuum conditions. Subsequently, the evacuated silica tube was inserted into a two-zone tubular furnace. The growth of CrSBr crystals took place over a period of 7 days, utilizing a temperature gradient ranging from 950 ${\degreecelsius}$ to 880 ${\degreecelsius}$, with a controlled heating and cooling rate of 1 ${\degreecelsius}$/min.
}

\vspace{3mm}

\noindent \small{\textbf{Device fabrication.}
Graphite, \textit{h}-BN, and CrSBr flakes were exfoliated using the scotch tape method in ambient conditions and transferred onto a Si substrate covered with a 285 nm thick SiO$_2$ layer. The heterostructures were then stacked using a pick-up and release method based on a poly propylene carbonate (PPC)/polydimethylsiloxane (PDMS) polymer stamp on a glass slide. Subsequently, the device underwent an acetone wash to remove PPC residual from the sample. A layer of poly methyl methacrylate (PMMA) was spin-coated and patterned using electron-beam lithography (EBL). The PMMA was developed using isopropyl alcohol, and the device was etched in a reaction-ion etching (RIE) system under O$_2$ and CHF$_3$ gas flow. Finally, Cr/Au electrodes, with a thickness of 5/35 nm were deposited on the device followed by a lift-off process to define metal contacts.
}

\vspace{3mm}

\noindent \small{\textbf{Electrical measurements.} 
Transport measurements were performed using a Heliox$^3$ He insert system equipped with a 14 T superconducting magnet, allowing temperature control down to 1.8 K. To analyze the $I$-$V$ characteristics of the tunneling barrier and investigate the MR, a Keithley 2636B source meter was used to apply the bias voltage, while the tunneling current was measured using a standard two-probe module. To ensure the detection of intrinsic signals and minimize the impact of Joule heating, the tunneling current was limited to several hundreds of nA, resulting in a total power of several hundreds of nW in a junction with an approximate area of 1 \textmu m$^2$.
}
\vspace{3mm}

\begin{acknowledgments}
This work was supported by the National Natural Science Foundation of China (No. 12425402 and No. 12250007), and the National Key R\&D Program of China (No. 2022YFA1203902). K.W. and T.T. acknowledge support from the JSPS KAKENHI (No. 21H05233 and No. 23H02052) and World Premier International Research Center Initiative (WPI), MEXT, Japan 

\end{acknowledgments}

\bibliographystyle{apsrev}
\bibliography{Reference_main}

\end{document}